# Reducing assembly complexity of microbial genomes with single-molecule sequencing

Sergey Koren[1, †], Gregory P Harhay[2], Timothy PL Smith[2], James L Bono[2], Dayna M Harhay[2], D. Scott Mcvey[3], Diana Radune[1], Nicholas H Bergman[1], and Adam M Phillippy[1]

Author Affiliations
1. National Biodefense Analysis and Countermeasures Center, 110 Thomas Johnson Drive, Frederick, MD 21702
2. USDA, ARS, U.S. Meat Animal Research Center, Clay Center, NE 68933
3. USDA, ARS, Center for Grain and Animal Health Research, Manhattan, KS  66502
† Corresponding author: korens@nbacc.net

# Abstract
**Background**
The short reads output by first- and second-generation DNA sequencing instruments cannot completely reconstruct microbial chromosomes. Therefore, most genomes have been left unfinished due to the significant resources required to manually close gaps in draft assemblies. Third-generation, single-molecule sequencing addresses this problem by greatly increasing sequencing read length, which simplifies the assembly problem.

**Results**
To measure the benefit of single-molecule sequencing on microbial genome assembly, we sequenced and assembled the genomes of six bacteria and analyzed the repeat complexity of 2,267 complete bacteria and archaea. Our results indicate that the majority of known bacterial and archaeal genomes can be assembled without gaps, at finished-grade quality, using a single PacBio RS sequencing library. These single-library assemblies are also more accurate than typical short-read assemblies and hybrid assemblies of short and long reads.

**Conclusions**
Automated assembly of long, single-molecule sequencing data reduces the cost of microbial finishing to $1,000 for most genomes, and future advances in this technology are expected to drive the cost lower. This is expected to increase the number of completed genomes, improve the quality of microbial genome databases, and enable high-fidelity, population-scale studies of pan-genomes and chromosomal organization.

# Background

As the cost of sequencing has dropped, the number of sequencing projects available in the GOLD database [1] has increased 4-fold from 2,905 in 2007 to 11,472 in 2011 [2]. However, many available genomes are heavily fragmented into hundreds or thousands of contigs, and many more are sequenced at low coverage and never submitted. This is in stark contrast to the era before the "next-generation" revolution, when many genomes underwent expensive manual gap-closing and sequence verification (finishing) before submission [3]. As sequencing costs have dropped, finishing has become impractical given the volume of sequencing data and manual effort required [4]. Only 32% of the genomes in the GOLD database are "complete" or "closed", meaning they contain no gaps. An even fewer number were "finished" by manually correcting errors and adding annotation [5]. This has hampered large-scale, structural analyses of bacterial genomes, and focused research instead on isolated genes and single-nucleotide polymorphisms (SNPs). While it remains impractical to manually finish all but the most important reference genomes, it is now possible to close microbial genomes at a reasonable cost using long-read, single-molecule sequencing and new assembly techniques [6-8]. This is expected to revitalize large-scale comparative genomic studies of whole genomes.

Single-molecule sequencing is a challenging problem that has not, until recently, resulted in a commercial product. Released in 2011, the PacBio RS is the only long-read, single-molecule sequencer currently available. In contrast to competing nanopore approaches [9-11], the PacBio RS utilizes an anchored polymerase and zero-mode waveguide to observe DNA polymerization in real time [12]. This instrument debuted as a rapid method for sequencing outbreak genomes [13, 14] and has since been demonstrated on eukaryotic genomes and transcriptomes [8]. Recent studies have focused on identifying DNA modification, such as methylation patterns, directly from the single-molecule sequencing data [15]. While adoption of this technology was initially slowed by the low accuracy of the single-pass sequences, recent advancements have demonstrated that this drawback can be algorithmically managed to produce assemblies of unmatched continuity [7, 8, 16]. Steady improvements to the PacBio technology continue to increase read lengths and yield [17], while future technologies promise to combine accuracy with length using either nanopores [11] or advanced sample preparation [18]. Improved microbial genome assembly is an obvious application of these recent developments in long-read sequencing.

Genome assembly is the process of reconstructing a genome from many shorter sequencing reads [19-21]. It is typically formulated as finding a traversal of a properly defined graph of reads, with the ultimate goal of reconstructing the original genome as faithfully as possible. Repeated sequence in the genome induces complexity in the graph and poses the greatest challenge to all assembly algorithms [22]. In addition, repeats are often the focus of analysis [23-25], making their correct assembly critical for subsequent studies. However, repeats can only be resolved by a spanning read or read pair that is uniquely anchored on both sides. Read pairs are typically used due to their length potential (tens of Kbp), but introduce additional complexity because they cannot be precisely sized. Alternatively, long-read sequencing promises to more accurately resolve repeats and directly assemble genomes into their constituent replicons. Figure 1 shows the benefit of increasing read length when assembling *Escherichia coli* K12 MG1655. This genome can only be assembled into a single contig when the read length exceeds the size of the longest repeat in the genome, a multi-copy rDNA operon. The rDNA operon, sized around 5–7Kbp, is the largest repeat class in most bacteria and archaea [26]. Therefore, sequencing reads longer than the rDNA operon, such as those produced by single-molecule sequencing, can automatically close most microbial genomes.

ALLPATHS-LG was the first assembler shown to produce complete microbial genomes using single-molecule sequences [7]. Utilizing a combination of PacBio RS single-molecule reads (2–3Kbp), short-range Illumina read pairs (<300bp insert), and long-range Illumina read pairs (3–10Kbp insert), ALLPATHS-LG assembles the Illumina reads first using a de Bruijn graph and incorporates PacBio reads afterwards to patch coverage gaps and resolve repeats. Riberio *et al.* tested this method on sixteen genomes and consensus accuracy was measured at 99.9999% on three genomes with an available reference. Four of the sixteen genomes were successfully assembled into a complete genome—the remaining genomes were all highly continuous but left unresolved due to large-scale repeats. These results are promising, especially in terms of consensus accuracy; however, the method requires two different sequencing platforms and three library preparations, which limits its efficiency. In addition, the jumping libraries were observed to be inconsistent at spanning large repeats due to biases in the library construction process.

Ideally, complete genomes could be reconstructed from a single fragment library, minimizing costs. Previously, pair libraries were the only sequencing method capable of spanning large repeats, such as the rDNA operon, but the PacBio RS is now capable of producing single-molecule reads of the same length. Leveraging this recent development, we present an approach for microbial genome closure that relies on overlapping and assembling single-molecule reads *de novo* rather than patching and resolving a short-read de Brujin graph. This exploits the log-normal sequence length distribution of the PacBio RS, which produces a significant fraction of sequences greater than 7Kbp [8]. These long reads can be utilized to span the longest repeat found in most microbial genomes, while the total coverage of reads can be used to construct a high-quality consensus. We estimate that this approach could automatically close >70% of the complete bacteria and archaea in GenBank, without the need for pair libraries, using the currently available PacBio "C2" chemistry. These single-library assemblies are also more accurate than typical short-read assemblies and hybrid assemblies of short and long reads. Finally, we show that the increased sequencing length of future technologies both decreases the coverage requirement and increases the number of genomes that can be closed using this method.

## Results and Discussion

Early single-molecule sequencing reads were too short and inaccurate to directly perform *de novo* assembly. Instead, it was shown that the reads could be corrected using a complementary technology prior to assembly [8, 14]. However, single-molecule read lengths have continued to improve, from a median length of 747bp in 2011 to 3,122bp in 2012 (Figure 2). Due to the increase in length, it is now possible to perform self-alignment and correction. This is because there are more detectable alignment seeds in a longer sequence versus a shorter sequence with equivalent error rate (Figure 3) [27, 28]. For example, 1.5Kbp sequences at 10% error are sufficient to reliably detect overlaps, but at 15% error, such as for XL-XL chemistry [29], 3.5Kbp sequences are required [27]. Based on this analysis, and the improving read length of the PacBio RS, we adapted the Celera Assembler PBcR pipeline [8] to support correction and assembly using only continuous long reads (CLR). This new version uses the BLASR software [27] to detect noisy overlaps between reads; an improved version of the PBcR algorithm to process overlaps and correct the long reads; and the Celera Assembler [30] for final assembly. The pipeline is designed to be compatible with future long-read data and has been tested on reads up to 64Kbp in length. The related HGAP software provided by PacBio [31] is a derivative of our correction and assembly pipeline [8] that also performs self-correction of CLR sequences followed by assembly with Celera Assembler. However, HGAP cannot use secondary sequencing data for correction. For consistent comparison to hybrid assembly, all reported results are from the PBcR version of the pipeline.

To evaluate the potential of long-read data to improve microbial genome assembly we first report the repeat complexity of all complete microbial genomes and predict the fraction that could be closed using a single PacBio sequencing library. We conclude with an analysis of six genomes sequenced using Illumina MiSeq, Roche 454, PacBio CCS, and PacBio CLR to evaluate performance on real data and compare PacBio CLR self-assembly versus a hybrid approach.

**Assembly complexity of NCBI genomes**
To describe the complexity of microbial genome assembly using long reads, we define three classes of microbial genomes in terms of repeat content and type, using the common rDNA operon as a benchmark (Figure 4). Class I genomes are defined as genomes with few repeats other than the rDNA operon. Class II genomes contain many mid-scale repeats, such as insertion sequences (IS) and simple sequence repeats (SSR), but the rDNA operon remains the largest. Class III genomes contain large phage-mediated repeats, segmental duplications, or large tandem arrays that are significantly larger than the rDNA operon. To delineate these classes, all maximal repeats greater than 500bp and 95% identity were identified for 2,267 finished microbial genomes available from NCBI. Figure 5 illustrates the distribution of maximum repeat size and repeat count for these genomes. The rDNA operon is clearly the largest repeat in most bacteria and archaea, with a tightly bounded size between 5–7Kbp. Thus, the threshold for Class III is set to a maximum repeat size of greater than 7Kbp. The boundary between Class I and II is less clear, and is set to 100 repeat copies for convenience.

Of the 2,267 genomes analyzed, 69.07%, 7.59%, and 23.33% comprise Class I, II, and III, respectively. It is important to note that Class I genomes can be assembled well, though not closed, by short-read sequencing, but Class II genomes, such as *Yersinia pestis*, have previously been considered the most difficult to assemble [32]. Now, with single-molecule sequencing reads in excess of 7Kbp, both the mid-range and rDNA repeats can be reliably spanned. This predicts automated closure of Class I and II genomes is now possible, and all but the longest Class III repeats can be resolved. We note that this analysis is database dependent and may underestimate the true membership of Class II and III, because repetitive genomes are the least likely to be complete. A table of repeat count and maximum repeat size for all complete genomes is provided as a supplementary file.

To evaluate the impact of long reads on microbial genome assembly, we simulated error-free sequences following PacBio C1, C2, XL-C2, XL-XL, and projected "ZL" corrected read length distributions. The hypothetical "ZL" chemistry is an extrapolation of annual PacBio chemistry improvements, and is based on a log-normal distribution with double the mean sequence length of the XL-XL chemistry. Sequencing coverage was generated from 50–200X for the same 2,267 genomes. For each genome, repeats were considered resolved if at least one simulated read spanned the entire repeat with unique anchors on both sides. In an overlap-based assembler, this is typically sufficient for resolving a repeat [33]. This estimates the potential resolving power of long reads in the absence of sequencing error. However, since corrected PacBio sequences achieve 99.9% accuracy in practice [8], it is also a reasonable approximation of true data.

Figure 6 shows the predicted number of genomes closed as well as the average number of assembly gaps for all variants of the simulated PacBio reads. For the C2 chemistry, 72.96% of the genomes contain zero gaps at 200X coverage, and the remaining genomes are well assembled but contain large, unresolved repeats. The expected number of gaps is 0.26±3.90, 0.34±1.39, and 2.89±2.92 for Class I, II, and III genomes, respectively. The benefit of additional C2 sequencing beyond 100X decreases rapidly, with almost no increase in the number of resolved genomes from 150X to 200X (34 additional genomes). Using the newer XL-C2 chemistry, the number of closed genomes plateaus much earlier, due to the larger number of long sequences. However, at least 50X is still required to generate an accurate consensus using only long-read sequencing

[31]. At 200X, upgrading to XL-C2 from C2 chemistry closes an additional 3.79% of genomes; upgrading from XL-C2 to XL-XL closes an additional 11.69%; and upgrading to "ZL" from XL-XL closes an additional 7%. There are diminishing returns in increasing read lengths, because many of the remaining unresolved repeats are more than double the average XL-XL lengths, requiring a jump in average read length to hundreds of kilobases to resolve.

No significant correlation between genome size and assembly continuity was found, which agrees with previous work that found no strong correlation between microbial genome size and repeat coverage [26]. However, assembly complexity is largely influenced by species-specific repeat structure. For example, for 200X C2, the expected number of gaps remaining in a *Bacillus anthracis* or *Yersinia pestis* strain is 0±0 and 0.42±0.51 respectively. However, some species, such as *Escherichia coli* (3.04±3.90), exhibit high variance due to strain-specific phage integrations, etc. An interactive Krona [34] chart detailing the expected number of gaps for all strains, species, genera, etc. under various coverage and chemistry scenarios is available as a supplementary file.

Based on these simulations, 150X is the recommended sequencing depth to maximize assembly continuity using the C2 chemistry. Given sequencing yields of approximately 300Mbp per SMRT cell after the RS II upgrade [35], this would require 3 cells for a 5Mbp genome. Due to the longer XL-C2 chemistry read length, only 100X of XL-C2 is required to maximize closure, or 2 cells for a 5Mbp genome. This equates to a lower cost versus previous approaches. The contract sequencing cost for a 5Mbp genome using the recipe of Riberio *et al.* [7] is ~$1,700, and can vary based on library preparation costs and multiplexing efficiencies (Methods). In contrast, the cost of our method, which relies on only a single library preparation, is ~$1,200 for 3 SMRT cells of C2, or ~$900 for 2 SMRT cells of XL-C2 (Methods). This represents a cost increase versus 100X of Illumina, which can be contracted for $300 or less (Methods), but the resulting Illumina assemblies are typically in hundreds of contigs and require heavy multiplexing to minimize costs [36, 37].

**Assembly and closure of real data**

To validate our approach, we sequenced the genomes of six bacteria of varying complexity and GC content (Table 1). For each genome, at least 200X of PacBio C2 CLR (continuous long reads) [12] was generated, along with 454 FLX+ [38], Illumina MiSeq [39], and PacBio CCS (circular consensus sequencing) for comparison to short-read and hybrid assemblies. For uniformity, datasets exceeding 200X CLR, 50X 454, 100X Illumina MiSeq, or 25X CCS were randomly down-sampled to these limits to reflect typical coverage depths. For the 5 novel genomes, a closely related genome was used to estimate assembly complexity. In the case of *E. coli* K12, the available reference sequence was used (Methods).

For long-read assembly, the genomes were then assembled using PacBio CLR in isolation and CLR corrected with the secondary technologies. The assemblies largely matched the simulated expectation, independent of the approach used (Table 2, chi-squared p-value < 0.015). The Class I genomes *Escherichia coli* K12, *Bibersteinia trehalosi*, and *Salmonella enterica* Newport, and the borderline Class III genome *Mannheimia haemolytica*, were all brought to closure using self-correction and at least one of the technology combinations. In nearly all cases, the assemblies of self-corrected CLR sequences outperformed the hybrid assemblies both in terms of continuity, error rate, and the assembly likelihood score. In contrast, assemblies of the hybrid data showed greater variability in performance, likely due to subtle errors introduced during correction. Only the self-corrected CLR assembly of *F. tularensis* did not match the simulated expectation. However, it is noted that this dataset has the lowest mean and median sequence length as well as a low fraction of sequences longer than 7Kbp relative to the other projects (8.52% versus an average of 13.47%). A machine failure occurred towards the end *F. tularensis* sequencing, possibly explaining the reduced performance of the preceding cells. An additional 100X coverage was generated for *F. tularensis*, bringing the assembly to expectation (Table 2).

For comparison to short-read sequencing and assembly, the 454 and MiSeq reads were assembled in isolation and the results compared to PacBio CLR assemblies (Table 3). As of writing this manuscript, the estimated sequencing costs for the assemblies presented in Table 3 are $300–$500 for 100X Illumina (paired HiSeq 2500 or MiSeq), $4,700 for 50X 454 (unpaired FLX+), and $1,400 for 200X PacBio (RS II C2). These estimates include library preparation and assume multiplexing efficiencies (Methods). Compared to the PacBio assemblies, the short-read assemblies were significantly less continuous, with well over 100 gaps and a 30-fold reduction in N50 contig size, on average. The cost to manually close these assemblies—estimated by Riberio *et al.* [7] to exceed $13,000—is an order of magnitude higher than any of the single-molecule methods (Methods). These results are consistent with expectation based on the short read lengths and repeat complexity of the presented genomes.

**Long-read assembly validation**
Reference-free assembly validation was performed on all assemblies using mapped Illumina MiSeq reads to estimate consensus error rates and determine an assembly likelihood score. For the majority of assemblies that did not utilize the MiSeq data, this represents an independent validation using a complementary data source. Consensus accuracy was estimated from single nucleotide discrepancies identified between the Illumina data and the assembly using FreeBayes [40], and assembly scores were computed using FRC$^{bam}$ [41], ALE [42], and LAP [43]. The latter two measure the likelihood of a set of Illumina reads being generated by the assembly—essentially how consistent the assembly is with the reads. FRC$^{bam}$, ALE, and LAP scored assemblies similarly, so Tables 2 and 3 list only the LAP scores and estimated consensus accuracy (a full validation report is provided as a supplementary file). In all cases self-correction produced the best LAP score with consistently high accuracies. The self-corrected assemblies averaged 99.9993% accuracy prior to polishing, and >99.99995% after re-alignment and polishing with Quiver. The second-generation assemblies averaged 99.9993% and 99.9992% accuracy for 454 and MiSeq, respectively. For comparison, a consensus accuracy of 99.999% (1 error in 100,000 bp) is considered finishing quality [4]. Thus, the automated assemblies of self-corrected PacBio CLR sequences surpass both the quality of second-generation assemblies and the quality standard for finished genomes.

A finished reference is available for *E. coli* K12 MG1655, which was one of the genomes sequenced here. However, the reference-free validation indicated the Quiver-polished assemblies of *E. coli* K12 were more consistent with the Illumina reads than the MG1655 reference sequence. The assemblies have both a better likelihood score and fewer single-nucleotide mapping discrepancies than the reference, suggesting that most differences between the assemblies and the published reference are true isolate-level variations. Table 4 reports these differences using the GAGE metrics [36].

Assembly performance varied depending on the correction method. On average, 454 correction was less accurate, which is unsurprising given the known homopolymer bias of this technology [38]. Sharing 454's length and error characteristics, a similar result would be expected if correcting with Ion Torrent data [44]. The CCS correction also underperformed the other methods, likely due to its lower per-read accuracy. Most promising, the self-corrected CLR sequences produced the fewest errors, even outperforming assemblies that included the Illumina data used for validation. This is consistent with the PacBio RS having low systematic bias, allowing a high-quality consensus to be generated even from low-accuracy reads [8]. Longer sequences can also be aligned more accurately, provided they are long enough to compensate for the error rate [28]. These results demonstrate that high-quality, high-continuity bacterial assemblies can now be generated using exclusively single-molecule sequencing data.

**Future technology**

All sequencing experiments above used the PacBio C2 chemistry released in February 2012. More recently, PacBio released the updated XL chemistry. Using *E. coli* K12 XL-C2 sequencing data provided by PacBio, we modeled the corrected read length distribution and simulated error-free sequences for the same 2,267 reference genomes as before. Using the longer sequences, more genomes are closed at lower coverage. This is due to the larger number of sequences over 7Kbp (22% for XL-C2 versus 16% with C2). Trading increasing read length for decreased accuracy can negatively impact alignment and assembly (e.g. XL-C2 versus XL-XL, Figure 3). For this reason, actual C2 and XL-C2 reads were found to assemble better than XL-XL in practice (data not shown). The RS II instrument upgrade increased throughput to 300Mbp per SMRT cell. Based on these numbers, two XL-C2 SMRT cells will be sufficient to close over 70% of known microbial genomes automatically, for less than $1,000 per genome. This includes the vast majority of Class I and II genomes, and predicts an average number of gaps of 2.93±2.90 for Class III genomes using XL-C2 sequencing. Similar predictions apply for any future technology capable of generating a significant throughput of reads above the golden 7Kbp threshold.

# Conclusions

Long, single-molecule reads are sufficient for the complete assembly of most known microbial genomes. The assemblies presented here have good likelihood and finished-grade consensus accuracy exceeding 99.9999%. By exploiting a model of the sequencing process, Quiver is able to improve assembly accuracy by 10QV, on average; and while there may be undiscovered biases in single-molecule sequencing, PacBio consensus accuracy always exceeded that of the second-generation sequencing data. In addition, assemblies of only single-molecule data consistently matched or exceeded the quality of both short-read and hybrid assemblies. Lastly, this approach requires only a single sequencing library, and reduces the time and cost of closure compared to previous approaches. However, for applications such as high-throughput SNP typing, draft Illumina sequencing is likely to remain the preferred option due to current throughput and cost advantages.

For Class III genomes that cannot be closed using current single-molecule sequencing, assembly continuity is significantly improved over first- and second-generation sequencing. In simulations, these most difficult genomes average only 2.89±2.92 gaps with C2 sequencing, suggesting that 99% of all known microbial genomes can be assembled into fewer than 10 contigs using currently available technology. This increase in continuity greatly reduces the required cost of manual closure, which directly correlates with the number of gaps. Complementary closure techniques, such as optical mapping [45], are also enhanced by longer contigs and can be used to efficiently close even the most complex genomes.

Long reads present a great opportunity, but also new challenges. For example, small replicons shorter than the typical 10Kbp library size may be inadvertently excluded from sequencing. We also noted low-abundance structural polymorphism in many of the samples analyzed. These mixed polymorphisms (e.g. inversions) would have been easily overlooked in fragmented assemblies of short reads. However, to fully capture such structural dynamics requires a graph-based representation of the genome, such as FastG [46] that allows for allelic diversity. In addition, long reads present algorithmic challenges to existing assemblers. Most existing assemblers are incapable of exploiting long reads. Celera Assembler and MIRA [47] are two exceptions, but these programs were developed for reads no more than 1Kbp and become cumbersome for very long reads. New algorithms, especially for consensus generation via multiple alignment, are needed to extract the full potential from these new data.

Finally, it is expected that future improvements to the PacBio chemistry, or the release of new technologies, will further extend the reach of microbial genome assembly. For example, the recent median read increase from PacBio C2 to XL chemistries (~2Kbp to ~3Kbp) is predicted to reduce the recommended coverage requirement by two fold. Thus, it is reasonable to expect that future chemistries with increased read lengths, and the corresponding throughput increases, will allow the full closure of most known bacteria and archaea at a cost of well under $1,000 a genome. This cost will continue to fall with future technology advances, improving reference database quality and enabling population-scale research on the structural dynamics of microbial genomes.

# Materials and Methods
**GOLD database and NCBI genomes**
To estimate the number of complete versus draft genomes, searches were performed on 3/4/2013 at Genomes OnLine Database (GOLD) [48]. The total number of projects with status "Complete and Published" was 2,427. The total number with status "Permanent Draft" was 1,752. The total number of projects of any status with sequencing status "Draft" was 3,389. A single project had project status "Complete and Published" and sequencing status "Draft" and 7 projects had project status "Permanent Draft" and sequencing status "Draft". These were excluded in the calculations to avoid double counting. The percentage of closed genomes was then computed as: 2427/(1752+2427+3389-1-7) = 2427/7560 = 32.10%.

Closed bacterial and archaeal genomes were obtained from NCBI [49] on 1/17/2013. This dataset contained 2,245 complete genomes (including constituent plasmids). The data was manually curated to remove 8 plasmid-only genomes and associate the loose plasmids with the appropriate genome based on identifiers listed in BioProject. Also, a total of 30 genomes combined more than a single assembly/BioProject and were separated. This resulted in the 2,267 (2,245 − 8 + 30) genomes used for analysis.

**Repeat analysis**
Genomic repeats were identified using Nucmer [50] `nucmer -maxmatch -nosimplify` and filtered using the associated delta-filter command `delta-filter -l500 -i95` to retain only repeats greater than 500bp and over 95% copy identity. Self-alignments on the main diagonal were discarded, and the repeat matches reduced to a set of intervals along the genome. Any interval contained within a larger interval was discarded; repeat count computed as the number of remaining intervals; and the largest interval noted as the maximum repeat size.

For each genome, error-free reads were uniformly sampled at 50-200X coverage and the read lengths were randomly chosen, with replacement, from real PBcR sequence distributions (*E. coli* K12 C1, C2 and XL-C2, and XL-XL chemistries) and a hypothetical future chemistry ("ZL"). A list of genomic repeats was compiled, as described above, and any abutting or overlapping repeats were merged. The simulated sequences were then mapped back to the genome using `nucmer -maxmatch`. A repeat was considered resolved if at least one read spanned the full length of the repeat with at least 40bp uniquely anchored on both sides. This reflects the default minimum overlap length needed by Celera Assembler to resolve a repeat. The simulation returns the expected number of contigs after breaking at any unspanned repeats.

For C1 chemistry, the length distribution was based on previously published results [8] using 50X CLR sequences and 100X Illumina for correction. For C2 chemistry, the length distribution was based on 200X of sequence corrected using 25X CCS. For XL-C2, the length distribution was based on 100X of sequence corrected using 25X CCS. For XL-XL chemistry, the length distribution was based on 50X distribution of match lengths after mapping sequences to the reference. The XL-XL sequences were later corrected and the observed corrected distribution closely matched the mapped distribution (mean 4,104.91 and 4,690.30 respectively, max 27,095 and 25,320, respectively). The "ZL" chemistry was based on a doubling of the XL-XL sequence length, similar to the past increase from C2 (mean = 2,476) to XL-XL (mean = 4,105). Log-normal distributions were fit to the data using the R function `rlnorm(100000, mean(log(values)), sd(log(values)))` and the maximum was limited to $\mu + 5 * \sigma$. The mean/sd values were (6.69, 0.37), (7.59, 0.67), (7.90, 0.63), and (8.02, 0.79) for C1, C2, XL-C2, and XL-XL distributions, respectively.

**PBcR correction pipeline**

Two notable improvements were applied to the PBcR algorithm [8]. One to improve detection of SMRTbell adapter sequences, and the other to fill coverage gaps introduced by the correction process. To remove SMRTbell adapters, short reads mapped to each long read are examined. If multiple short reads match in both a forward and reverse orientation around a common point, the long read is split at this position. To identify sequences with multiple breakpoints, the above procedure can be applied recursively to the split subsequences. To patch an alignment coverage gap, short reads surrounding the gap are first identified. All other long reads containing these short reads in the same order and orientation are recruited and the consensus of the long read sequences is used to fill the gap. Only the surrounding short reads within a fixed window are used for recruitment. On an *E. coli* K12 test dataset, the average corrected read length increased to 4,187 from 2,493 when this feature was enabled while maintaining a corrected read identity above 99.6%. The assembly N50 also increased from 3.32Mbp to 4.65Mbp.

It was previously assumed that correction using only PacBio CLR reads was not feasible [8]. However, this analysis was based on the C1 chemistry with a median read length of only 540bp and error rate of 16.3%. We reproduced the analysis in Chaisson *et al.* [27] on varying read lengths and error rates. Using C1 and C2 sequencing data, we compared the length and identity of the overlaps when mapping sequences to themselves versus mapping them to the reference. The predicted overlaps closely match the expected overlaps for alignments of C2 reads. However, C1 overlaps are under-detected because they are not sufficiently long to compensate for their error rate. Thus, self-correction is not feasible with the C1 chemistry, but more recent chemistries (C2 onward) allow self-correction due to the increased read length.

**Sequence generation**

Libraries were prepared for each bacterial strain using kits provided by the manufacturer of the sequencing platform, as suggested by the product manuals. PacBio CLR reads were generated from libraries made with genomic DNA sheared to average 8-10 kb using either Hydroshear (Digilab; Marlborough, MA), or Covaris G-tube (Covaris, Inc., Woburn, MA). SMRTbell libraries were prepared from these fragments, and bound to eC2, C2, or XL versions of polymerase as suggested by the manufacturer. For eC2 and C2 polymerases, bound complexes were passively loaded into the SMRT cells on the instrument. For XL polymerase, bound complexes were adhered to MagBeads as recommended and actively loaded. At the time of this data collection, the stage start option for sequence collection was not available, so the default mode of data collection was used. PacBio CCS reads were generated from libraries made with genomic DNA sheared to average 300-800 bases using a Covaris S220 instrument according to the instrument recommendations for these fragment sizes. The libraries were bound to C2 polymerase, and passively loaded into the SMRT cells for sequencing.

Illumina libraries were prepared using TruSeq DNA sample prep kits (Illumina Inc., San Diego, CA) as recommended by the included instructions. DNA was sheared to approximately 500–800 bases prior to library construction using a Covaris S220 instrument. The libraries were sequenced using 2x150 or 2x250 paired end protocols on a MiSeq instrument (Illumina), as recommended by the manufacturer.

Libraries for sequencing on the GS FLX+ platform (454 Life Sciences, Branford CT) were prepared with GS Rapid Library Prep Kits as suggested by the manufacturer. Genomic DNA was sheared to approximately 2Kbp prior to library construction using a Covaris S220, and sequenced on the instrument using recommended emPCR and sequencing conditions for GS FLX+ sequencing kits.

**Long-read correction and assembly**
For each genome, 200X of PacBio CLR sequences were corrected using pacBioToCA. In addition to self-correction, hybrid correction using 100X MiSeq, 50X 454, and 25X CCS was performed. Whenever more data was available, it was randomly down-sampled to these values for consistency. All runs used the same default parameters with the exception of the genome size, which was manually approximated beforehand for each genome. After correction using CCS, MiSeq, or 454, the corrected sequences were trimmed by quality and subsampled before assembly. The trimming procedure is an automated step that selects, via dynamic programming, the largest range for each corrected read such that the average consensus quality score exceeds QV 54.5. The minimum overlap length was adjusted based on the average length of the trimmed sequences. After trimming, only the longest 25X of the corrected reads were assembled. Overlap-based assemblers tend to underperform on high coverage data [19, 21], so this sampling step both reduces runtime and helps improve assembly continuity. After assembly, contigs with fewer than 100 reads were discarded and the rest polished following PacBio's guidelines. The assembly was imported as the reference and 8 SMRT cells (5 for FT) were aligned to the genome using the `RS_Resequencing` pipeline. SMRTanalysis 1.4.0 was run as `smrtpipe.py –params=settings.xml xml:input.xml`. Any lower-case bases or ambiguities (Ns) remaining were trimmed from the beginnings and ends of contigs.

**Secondary sequencing correction and assembly**
For each genome, the same data that was used for correcting PacBio RS sequences was also assembled in isolation. At most 50X of 454 data was assembled using Newbler v2.8 [38]. Newbler ran as `runAssembly –o asm –cpu 8 *.sff`. At most 100X of Illumina MiSeq data was assembled using SPAdes v2.5.0 [51] and MaSuRCA v1.9.5 [52] as these were the top-performing assemblers in the GAGE-B competition [37]. SPAdes ran as `spades.py -k 21,33,55,77 --careful --pe1-1 miseq.1.fastq --pe1-2 miseq.2.fastq -o spades.` MaSuRCA ran as `runSRCA.pl config.txt` followed by `bash assemble.sh`. The contents of the config.txt file were:
```
PATHS
JELLYFISH_PATH=/full/path/to/MSR-CA/bin
SR_PATH=/full/path/to/MSR-CA/bin
CA_PATH=/full/path/to/MSR-CA/CA/Linux-amd64/bin
END

DATA
PE= pe 500 200 miseq.1.fastq miseq.2.fastq
END

PARAMETERS
GRAPH_KMER_SIZE=auto
```

```
USE_LINKING_MATES=1
LIMIT_JUMP_COVERAGE=60
CA_PARAMETERS=ovlMerSize=30 cgwErrorRate=0.25 ovlMemory=4GB
KMER_COUNT_THRESHOLD=1
NUM_THREADS=16
JF_SIZE=5000000000
DO_HOMOPOLYMER_TRIM=0
END
```
Both SPAdes and MaSuRCA assemblies were polished using iCORN [53]. The 454 assembly along with the four Illumina MiSeq assemblies were validated as described below. Only the best scoring Illumina assembly for each genome is included in Table 3 (a full validation report for all generated assemblies is provided as a supplementary file).

**Validation**

For *E. coli* K12, the MG 1655 reference [GenBank:NC000913] is available. Reference-based validation was performed using the GAGE scripts and metrics [36]. The SNP count between references and assemblies was calculated using Nucmer and show-snps [50]. For all genomes, reference-free validation was also performed. When no reference was available, a near-neighbor was included for comparison in validation results. For *E. coli* O157:H7 the reference *E. coli* O157:H7 Sakai [GenBank:NC_002127], [GenBank:NC_002128], [GenBank:NC_002695] was used. For *F. tularensis,* the reference *F. tularensis* subsp. Holarctica OSU18 [GenBank:NC_008369] was used. For *S. enterica* Newport, the reference *S. enterica* Newport SL254 [GenBank:NC_011079], [GenBank:NC_011080], [GenBank:NC_009140] was used. For *S. enterica* Newport, Illumina reads and contigs corresponding to two short plasmids (<10Kbp) and the phiX control were removed prior to validation. Reference-free validation was performed with FRC$^{bam}$ [41] ALE [42], and LAP [43]. These tools require paired-end sequences for validation, so Illumina data was mapped to the assemblies. For FRC, the bowtie command `bowtie –I <min> -X <max> -f –l 25 –e 140 –best –k 1 –S` was used, based on the example provided with the source. For ALE, the bowtie command `bowtie –I <min> -X <max> -f –l 10 –e 300 –a –v 1 –S` was used. LAP has a built-in bowtie2 procedure that was used. To call SNPs, a random 100X subset of reads were selected from the mapped, left end of Illumina pairs. Left ends were selected as they were observed to be higher quality and a larger fraction mapped to the assemblies. From these reads, SNPs and INDELs were called using FreeBayes. FreeBayes was run with the command `freebayes –C 2 –0 –O –q 20 –F 0.51 –z 0.02 –X –U –p 1`.

**Sequencing Cost Estimate**

Sequencing costs for PacBio RS and Illumina library preparation were taken on 07/16/2013 from the Duke Genome Sequencing & Analysis Core Resource [54]. The price for an external domestic (U.S.A.) institution was used. The throughput of a single PacBio RS cell was assumed to be 125Mbp for CLR and 20Mbp for CCS. The throughput of a single PacBio RS II cell was assumed to be 300Mbp. The throughput of a GS FLX+ was assumed to be 700Mbp. The throughput of a MiSeq run was assumed to be 3Gbp. The throughput of a HiSeq flow cell was assumed to be 300Gbp (37.5Gbp per lane). Library costs are listed as advertised by Duke, but could potentially be lowered for large-scale projects via automation.

The cost of multiplex sequencing on a HiSeq 2500 was computed as:
100X Illumina Paired-end = Library prep + 1/75 HiSeq lane = $283.05 + $39.44 = $322.49

The cost of multiplex sequencing on a MiSeq was computed as:
100X Illumina Paired-end = Library prep + 1/6 MiSeq Run = $283.05 + $196.55= $479.60

The cost of half-plate of sequencing on a 454 GS FLX+ was computed as:

50X 454 Fragment = Library prep + 1/2 Plate = $834.99 + $3,839.84 = $4,674.83

The cost of the ALLPATHS-LG [7] recipe was computed as:
50X Illumina Paired-end = Library prep + 1/12 MiSeq Run = $283.05 + $98.28 = $381.33
50X Illumina Mate-pair = Library prep + 1/12 MiSeq Run = $667.03 + $98.28 = $765.31
50X PacBio RS II = Library prep + 1 SMRT cell = $365.05 + $252.22 = $617.27
Total = $1,763.91 ($1,628.98 using 96-way multiplexed HiSeq).
This is consistent with the estimate of $1,669 for Illumina sequencing and $1,365 for PacBio RS sequencing (total $3,034) given in the supplementary materials of Riberio *et al.* [7], adjusted for cost decreases over the past year.

The cost of PacBio RS sequencing was computed as:
25X PacBio RS CCS = Library prep + 6 SMRT cells = $365.05 + $1,513.32 = $1,878.37
150X PacBio RS C2 = Library prep + 6 SMRT cells = $418.68 + $1,513.32 = $1,932.00

Given the recent PacBio RS II update, the per-cell yields have increased from 125Mbp to 300Mbp. Thus, 2 SMRT cells are sufficient for 100X coverage and 3 SMRT cells are sufficient for 150X coverage. The cost of RS II sequencing was computed as:
200X PacBio RS II C2 = Library Prep + 4 SMRT cells = $418.68 + $1,008.88 = $1,427.56
150X PacBio RS II C2 = Library Prep + 3 SMRT cells = $418.68 + $756.66 = $1,175.34
100X PacBio RS II XL = Library Prep + 2 SMRT cells = $418.68 + $504.44 = $923.12

If yields improve further to 500Mbp, 1 SMRT cell would become sufficient for 100X of XL-C2, at a cost of:
100X PacBio RS (proj.) XL-C2 = Library Prep + 1 SMRT cell = $418.68 + $252.22 = $670.90

For comparison, Riberio *et al.* [7] estimates the cost of closure at $13,124 assuming Illumina paired-end and jumping library preps and a resulting assembly with 50 gaps. Based on our simulation of NCBI genomes, if it is assumed all repeats below 500bp are resolved, the average number of gaps per genome is 46±52, with a maximum of over 500 gaps.

**Data Release**

All data, supplementary files, assemblies, and code described here are available at:
http://www.cbcb.umd.edu/software/PBcR/closure/index.html

The sequencing data and assemblies for novel strains generated for this study have been deposited in NCBI under the following accessions: *E. coli* K12 [PRJNA194437], *B. trehalosi* [PRJNA157929], *M. haemolytica* [PRJNA212438], *S. enterica* [PRJNA51643], *E. coli* O157:H7 [PRJNA63279], and *F. tularensis* [PRJNA212941]. The Illumina MiSeq *E. coli* K12 sequencing data is available from the Illumina Scientific Data Website (http://www.illumina.com/systems/miseq/scientific_data.ilmn).

## List of Abreviations
bp - base pair
Kbp - thousand base pairs
Mbp - million base pairs
CA – Celera Assembler, a *de novo* whole genome shotgun assembler originally designed for Sanger [55] sequences and since adapted to 454 [30] and PacBio RS [8] sequences.

CCS – PacBio RS Circular Consensus Reads, sequences generated from a short library prep. By reading each short molecule multiple times, the error rate is reduced, improving accuracy to over 99%.

CLR – PacBio RS Continuous Long Reads, sequences generated from long molecules. Individual sequences have low accuracy, in the 85-88% range.

INDEL – insertion/deletion, the removal or addition of at least one base pair to the assembly when compared to either the reference or sequencing data.

PBcR – PacBio corrected Reads, CLR sequences after correction by pacBioToCA using either the CLR reads themselves or a complementary high-identity technology.

QV – "Phred" style consensus quality value, computed from the mean number of bases between assembly errors. Can be converted to an error probability $P=10^{-QV/10}$.

SNP – Single Nucleotide Polymorphism, the change of a single base pair in the assembly as compared to the reference or the sequencing data.

# Competing Interests

None declared.

# Authors' Contributions

SK and AMP conceived the project. SK implemented the algorithms and ran experiments. SK and AMP wrote the manuscript and generated figures. GPH ran assembly experiments and validated sequencing. DR performed manual validation of assemblies. DH, JB, DSM, NHB, and TS isolated strains and provided sequencing. All authors read and approved the final manuscript.

# Description of Additional Files

ValidationResults.xlsx – Reference-free assembly validation. The columns are as described in Table 2 with the addition of the max contig, unmated LAP score, FRC feature count, and ALE scores.

RepeatStats.xlsx – Listing of repeat count and maximum repeat size for all complete genomes.

report.log.krona.html – An interactive Krona [34] chart detailing the expected number of gaps for all strains, species, genera, etc. under various coverage and chemistry scenarios.

# Acknowledgements

The authors thank Renee A. Godtel and Robert T. Lee for generating CLR, 454, and Illumina sequence data. We thank the NBACC Genomics group for generating CLR and 454 data. We thank Illumina Inc. for providing MiSeq sequencing data. We also thank Todd J. Treangen and Brian Ondov for their careful reading of the draft and valuable discussions. We thank Pacific Biosciences for providing *E. coli* K12 sequencing data, Aaron Klammer, Jason Chin, David Alexander, and Edwin Hauw in particular for useful assistance and discussions. We thank David A. Rasko for providing *E. coli* K12 MG1655 DNA. We thank Mick Watson for feedback on a pre-print version of the manuscript.

The contributions of SK, DR, NHB, and AMP were funded under Agreement No. HSHQDC-07-C-00020 awarded by the Department of Homeland Security (DHS) for the management and operation of the National Biodefense Analysis and Countermeasures Center (NBACC), a Federally Funded Research and Development Center. The views and conclusions contained in The contributions of GPH, JB, DSM, DH, TPLS were funded by United States Department of Agriculture, Agricultural Research Service. The contributions of DSM were funded by the Nebraska Agriculture Experiment Station (USDA Formula Funds and UNL funds) and the Nebraska Veterinary Diagnostic Laboratory. Genome sequencing of the *E. coli* O157:H7 and *S. enterica* Newport strains was supported in part by The Beef Checkoff (JLB & DBH).

# Figure Captions

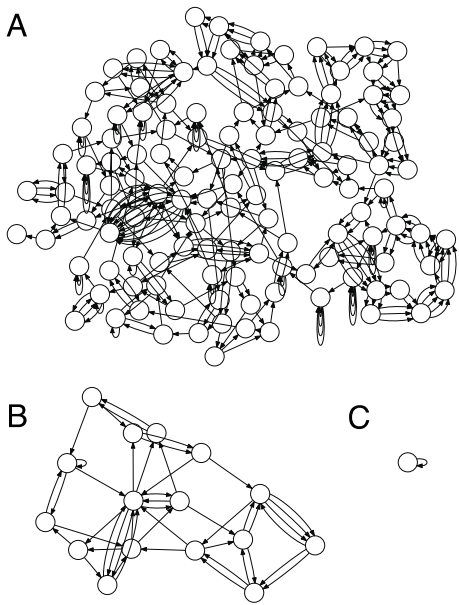

**Figure 1 – Genome assembly graph complexity is reduced as sequence length increases.** Three de Bruijn graphs for *E. coli* K12 are shown for *k* of 50, 1,000, and 5,000. The graphs are constructed from the reference and are error-free following the methodology of Kingsford *et al.* [32]. Non-branching paths have been collapsed, so each node can be thought of as a contig with edges indicating adjacency relationships that cannot be resolved, leaving a repeat-induced gap in the assembly. A) At *k*=50, the graph is tangled with hundreds of contigs. B) Increasing the k-mer size to *k*=1,000 significantly simplifies the graph, but unresolved repeats remain. C) At k=5,000, the graph is fully resolved into a single contig. The single contig is self-adjacent, reflecting the circular chromosome of the bacterium.

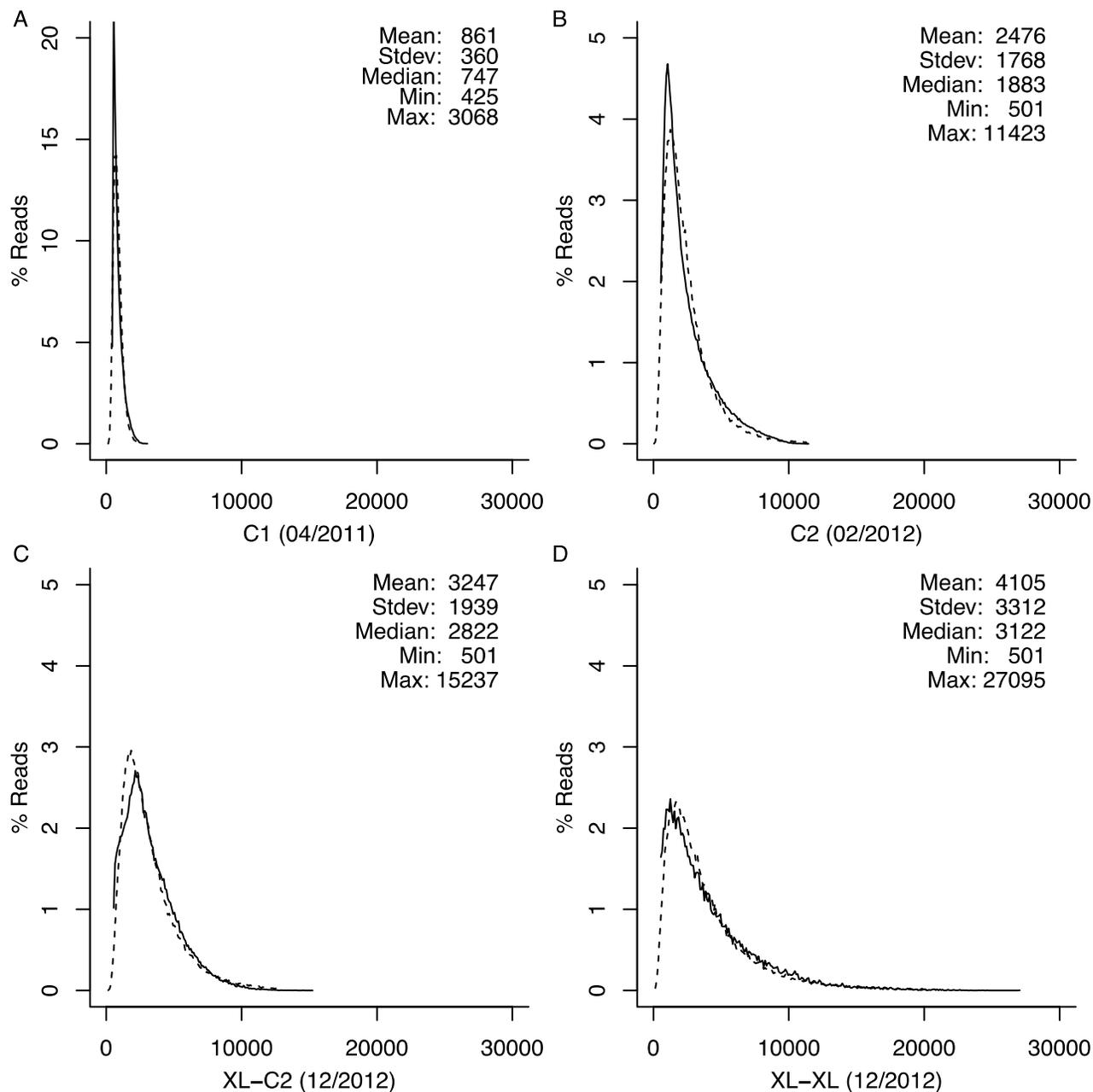

**Figure 2 – Improving PacBio RS sequence lengths.** The sequence length histograms of four PacBio RS chemistries are shown using 100bp buckets. Solid lines correspond to observed sequence lengths and dashed lines correspond to fitted log-normal distributions [17] with the specified mean and standard deviation. Since the initial instrument release in April 2011, the average sequence length increased over 3.5-fold through December 2012. A) The original C1 chemistry, released in April 2011; B) C2 chemistry, released in February, 2012; C) XL-C2 chemistry, released in December 2012; and D) XL-XL chemistry, released in December 2012.

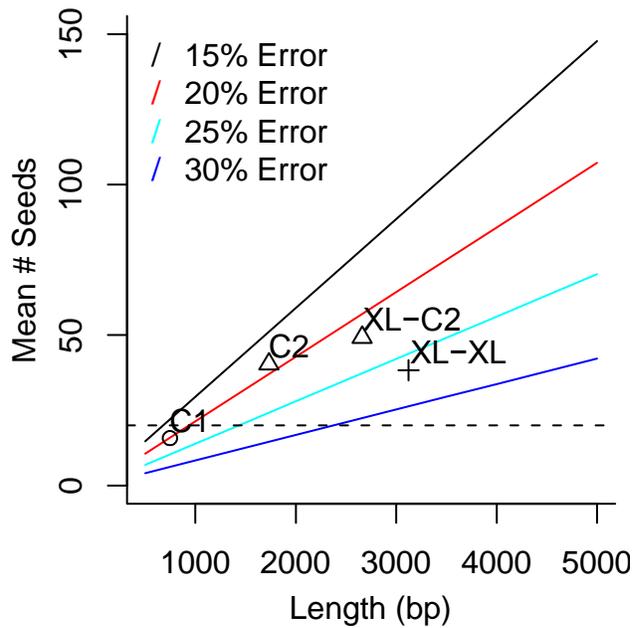

**Figure 3 – Sequence length compensates for increased error.** The mean number of expected 10bp seeds (the default in BLASR) was computed for each sequence length and error rate following the method in Chaisson *et al.* [27]. Additional seeds decrease the number of matches that have to be examined, decreasing runtime and increasing accuracy. For example, increasing the number of 15bp seeds from 10 to 20 reduces the number of sequences with over 100 matches to the human reference by 25% [27]. Points correspond to the median sequence length and observed error rate of four PacBio RS sequencing chemistries. Sequence lengths also compensate for increased error since more seeds can be found in a longer sequence. For example, 20 seeds (dashed line), can be found both in a 0.75Kbp sequence at 15% error and a ~2.5Kbp sequence at 30% error.

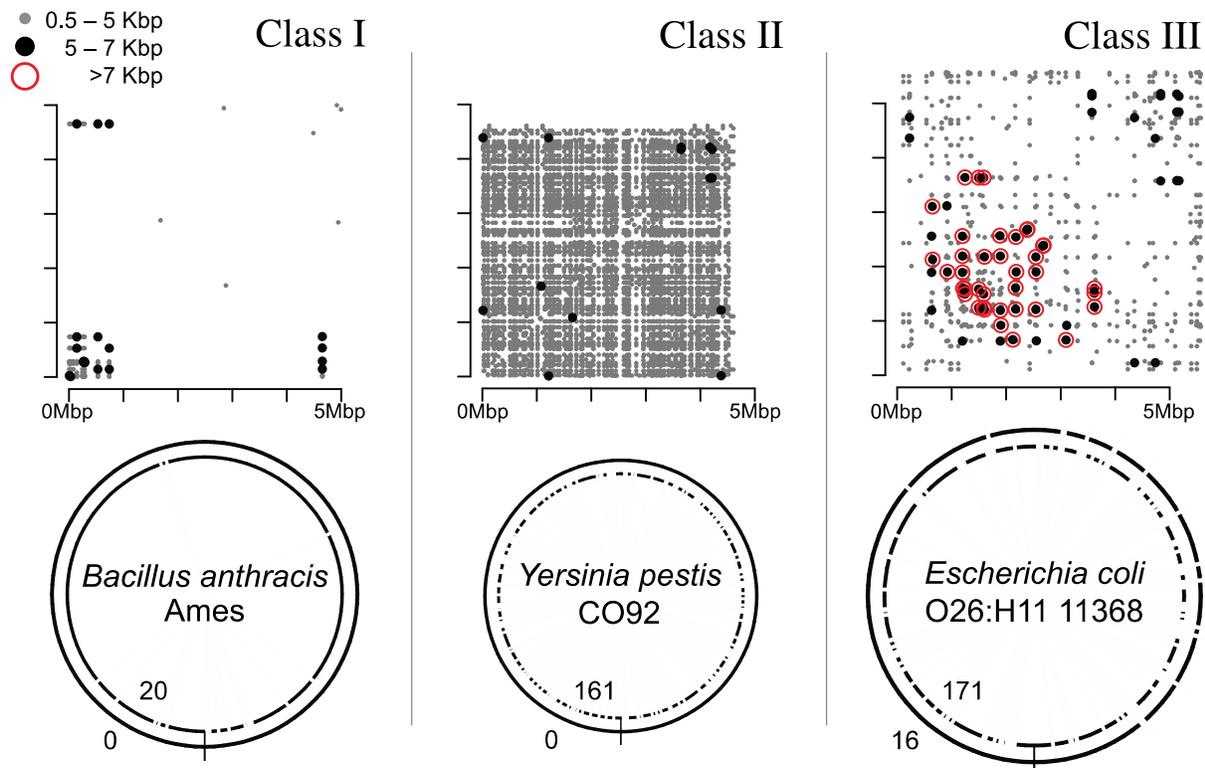

**Figure 4 – Three classes of microbial genome assembly complexity.** The top row illustrates repeat content via an alignment dotplot in *Bacillus anthracis* Ames, *Yersinia pestis* CO92, and *Escherichia coli* O26:H11 11368. For a repeat occurring at two distinct positions *x* and *y* in the genome, a dot of the corresponding size is placed on the matrix at [*x*,*y*]. The bottom row illustrates assemblies of these genomes using 200X simulated PacBio C2 sequencing (outer circle), and infinite coverage of 500bp, perfect reads (inner circle). The number of gaps in each assembly is noted. Class I genomes have few repeats except for the rDNA operon sized 5–7Kbp. In this case, both short reads and PacBio reads can generate a continuous assembly. Class II genomes have many repeats, such as IS elements, but none greater than 7Kbp. In this case, the PacBio reads can completely assemble the genome, while the short-read assembly is heavily fragmented. Class III genomes contain large, often phage-related, repeats >7Kbp. In this case, no technology can generate a complete genome. However, the PacBio assembly is significantly more continuous than the short-read assembly.

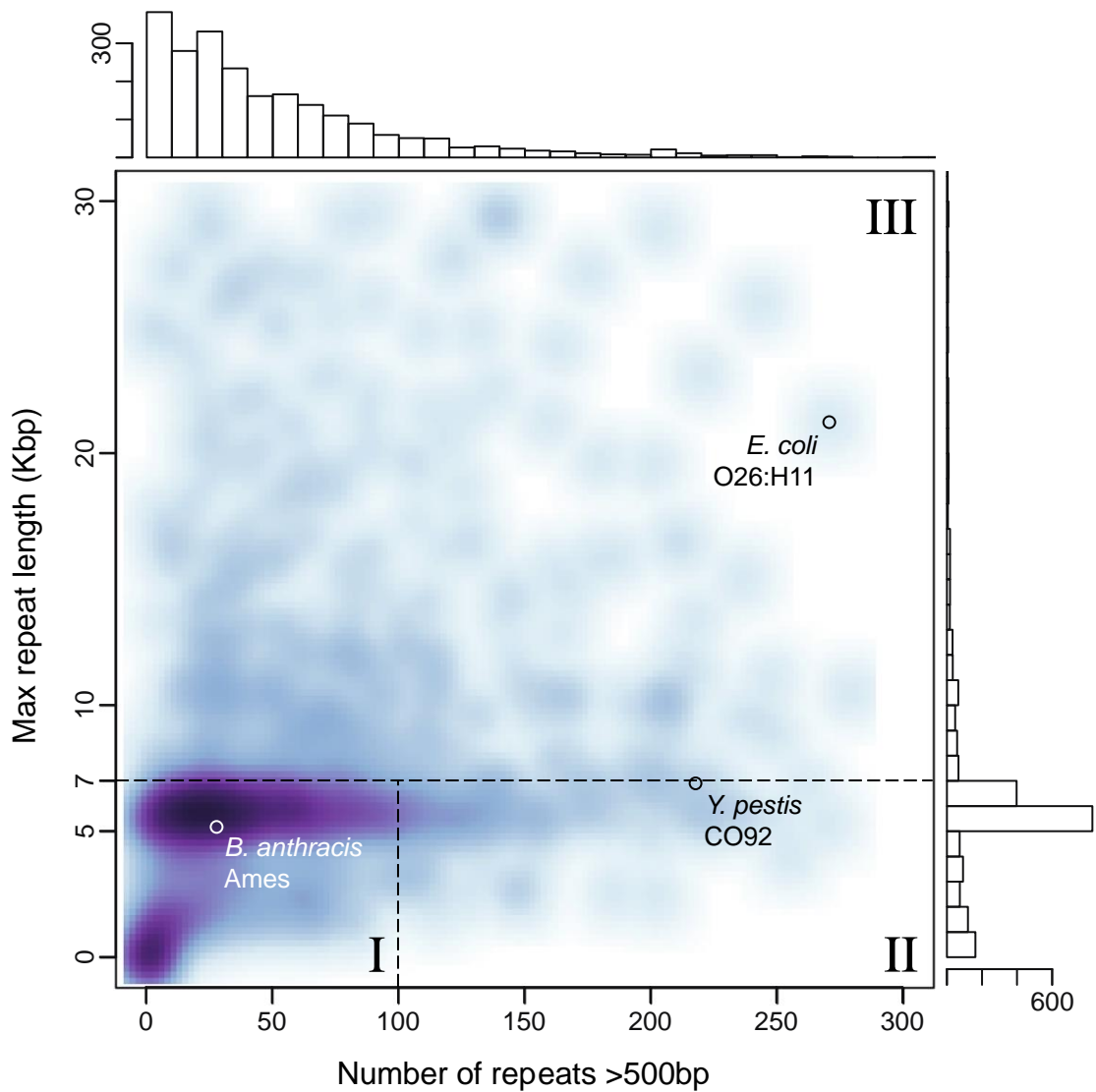

**Figure 5 – Repeat count versus maximum repeat length for 2,267 complete genomes.** For each genome, the number of repeat regions >500bp is given on the horizontal axis and the size of the largest repeat in the genome is given on the vertical axis. A smoothed scatterplot of all complete genomes is given center, with the corresponding histograms for each axis given on the top and right. The figure is cropped to show only repeat counts <300 and maximum repeat size <30Kbp. This comprises 95% of the data, with the remaining 5% containing a maximum repeat >30Kbp or more than 300 repeats. In the extremes, Class II genomes can reach over 800 repeat copies, and Class III genome repeats can exceed 100Kbp [26, 56].

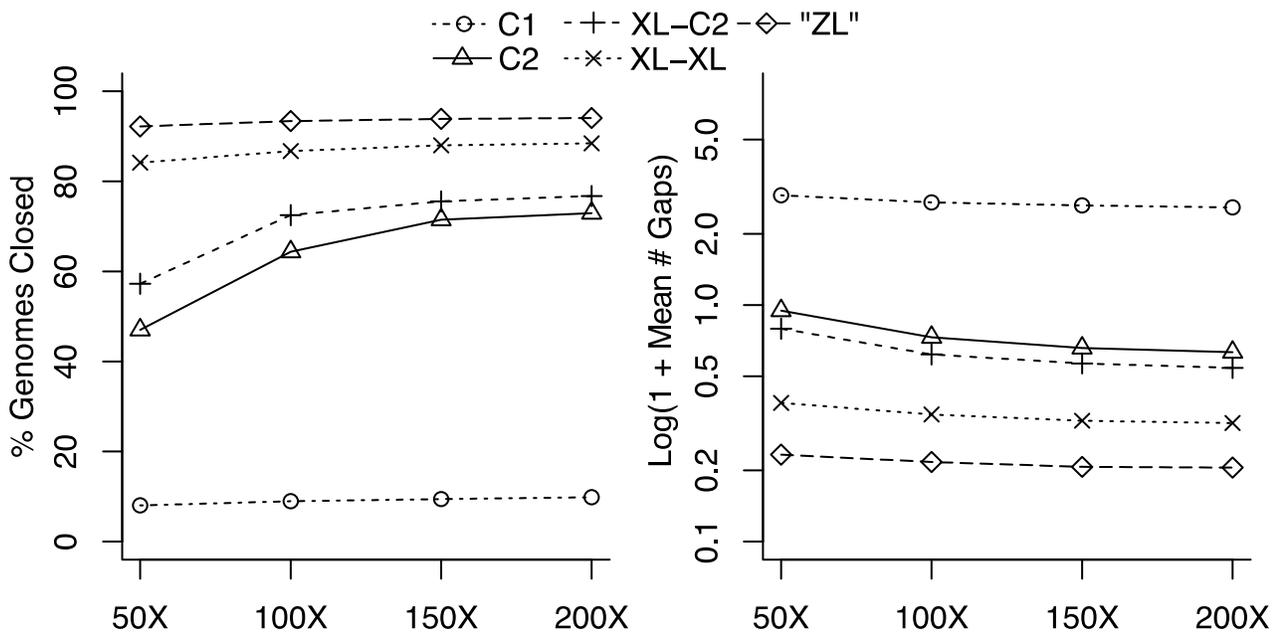

**Figure 6 – Assembly improvement with increasing coverage and read length.** Simulated assembly results on all complete NCBI references as of January 2013 using PacBio RS C1, C2, XL-C2, XL-XL, and projected "ZL" chemistries. The two figures show the percentage of genomes closed (left) and the average number of remaining gaps (right) with increasing sequencing coverage. C2 and newer chemistries can span the rDNA repeat and thus close many more genomes than the C1 chemistry. However, beyond 150X C2 there is limited benefit from further sequencing because the remaining repeats are too long to resolve. The longer chemistries saturate most repeats and gain little benefit from additional coverage over 50X. Resolving the remaining repeats requires a jump in sequence length to hundreds of kilobases.

# Tables

**Table 1: Total sequence and estimated coverage for six sequenced genomes.** The table lists available sequencing coverage (in base pairs and estimated fold coverage) of six bacterial genomes chosen to validate the closure recipe, for the four types of data produced (PacBio CLR, PacBio CCS, 454 titanium, and MiSeq paired end). Sequencing data was randomly down sampled following the procedure described in methods. For each genome, the expected genome size, GC content, and complexity class is included. PB% Error (200X): The estimated error rate based on BLASR [27] mappings of the analyzed random 200X subset to the assembly. PB % > 7Kbp (200X): The percentage of bases in sequences longer than 7Kbp, before correction. PB Coverage >7Kbp (200X): The coverage represented by sequences longer than 7Kbp, before correction.

| Organism | Genome Mbp | GC% | Class | PB CLR Gbp | PB % Error (200X) | PB % >7Kbp (200X) | PB Coverage >7Kbp (200X) | PB CCS Gbp | 454 Gbp | MiSeq Gbp |
|---|---|---|---|---|---|---|---|---|---|---|
| *Escherichia coli* K12 MG1655 | 4.65 | 50% | I | 1.4 (294X) | 10.44 | 16.07 | 32.11X | 0.2 (44X) | 0.19 (42X) | 1.73 (372X) |
| *Escherichia coli* O157:H7 F8092B [57] | 5.52 | 50% | III | 2.2 (397X) | 10.49 | 13.37 | 26.75X | N/A | 0.22 (40X) | 0.65 (118X) |
| *Bibersteinia trehalosi* USDA-ARS-USMARC-192 | 2.41 | 41% | I | 1.1 (439X) | 13.17 | 5.56 | 11.12X | 0.09 (35X) | 0.15 (59X) | 0.62 (247X) |
| *Mannheimia haemolytica* USDA-ARS-USMARC-2286 | 2.55 | 41% | III | 0.5 (212X) | 10.46 | 23.06 | 47.12X | 0.1 (42X) | N/A | 0.39 (85X) |
| *Francisella tularensis* 99A-2628 | 1.90 | 32% | III | 0.7 (331X) | 11.82 | 8.52 | 17.03X | N/A | 0.09 (40X) | 11.73 (5870X) |
| *Salmonella enterica* Newport SN31241 [58] | 5.01 | 52% | I | 1.1 (217X) | 12.33 | 9.30 | 18.61X | 0.1 (22X) | 0.13 (25X) | 0.28 (56X) |

**Table 2: Genome assembly continuity and correctness using hybrid and self-correction approaches.** Organism: The genome being assembled. Corrected By: the short-read data used for correction. Assembly bp: the total number of base pairs in all contigs (only contigs containing at least 100 reads are included in all results). # Contigs (expected): predicted number of contigs for a known reference (or near-neighbor). # Contigs (actual): The number of contigs comprising the assembly. N50: $N$ such that 50% of the genome is contained in contigs of length $\geq N$. LAP: The assembly likelihood score. A score closer to zero indicates a better assembly. # Discordant Bases: The number of SNPs and indels identified by mapping MiSeq sequences back to the assembly and recording discrepancies. Each incorrect base is counted (i.e. an indel that is a deletion of two bases from the assembly counts as two in this column). QV: estimated from the number of discordant bases as $\log_{10}\left(\frac{\text{assembly length}}{\text{\# incorrect bases}}\right) * 10$. Assemblies were generated by Celera Assembler [30] followed by post-processing with Quiver [31].

| Organism | Corrected By | Assembly bp | # Contigs (expected) | # Contigs (actual) | N50 (expected) | N50 (actual) | LAP | # Discordant Bases | QV |
|---|---|---|---|---|---|---|---|---|---|
| *E. coli* K12 | Reference | 4639675 | | 1 | 4639675 | N/A | -9.65E+07 | 4 | >60 |
| | MiSeq 100X | 4647253 | 1 | 2 | | 2367319 | -9.64E+07 | 3 | >60 |
| | 454 50X | 4649004 | 1 | 1 | | 4649004 | -9.64E+07 | 3 | >60 |
| | CCS 25X | 4653267 | 1 | 1 | | 4653267 | -9.64E+07 | 3 | >60 |
| | **Self** | **4653486** | 1 | **1** | | **4653486** | **-9.64E+07** | **3** | **>60** |
| *E. coli* O157:H7 | Near Neighbor | 5594477 | | 3 | 3776951 | N/A | -3.82E+07 | 1282 | 36.40 |
| | MiSeq 100X | 5624394 | 10 | 10 | | 3089011 | -3.66E+07 | 4 | >60 |
| | 454 40X | 5613057 | 10 | 12 | | 927294 | -3.67E+07 | 13 | 56.35 |
| | **Self** | **5611389** | 10 | **9** | | **4324437** | **-3.66E+07** | **0** | **>60** |
| *B. trehalosi* | MiSeq 100X | 2402545 | | 6 | | 1603511 | -3.28E+07 | 1 | >60 |
| | 454 50X | 2413761 | | 4 | | 1051672 | -3.27E+07 | 2 | >60 |
| | CCS 25X | 2411501 | | 1 | | 2411501 | -3.27E+07 | 0 | >60 |
| | **Self** | **2411068** | | **1** | | **2411068** | **-3.27E+07** | **0** | **>60** |
| *M. haemolytica* | MiSeq 100X | 2712467 | | 1 | | 2712467 | -3.31E+07 | 0 | >60 |
| | CCS 25X | 2739949 | | 2 | | 2686992 | -3.31E+07 | 0 | >60 |
| | **Self** | **2736037** | | **1** | | **2736037** | **-3.31E+07** | **0** | **>60** |
| *F. tularensis* | Near Neighbor | 1895727 | | 1 | 965253 | N/A | -1.33E+07 | 113 | 42.25 |
| | MiSeq 100X | 1879071 | 3 | 10 | | 357518 | -1.33E+07 | 0 | >60 |
| | 454 50X | 1863947 | 3 | 15 | | 201203 | -1.33E+07 | 0 | >60 |
| | Self | 1828135 | 3 | 8 | | 401731 | -1.33E+07 | 0 | >60 |
| | **Self (300X)** | **1877407** | 3 | **3** | | **573021** | **-1.33E+07** | **0** | **>60** |
| *S. enterica* Newport | Near Neighbor | 5007719 | | 2 | 4827641 | N/A | -2.26E+07 | 20 | 53.99 |
| | MiSeq 56X | 5027784 | 4 | 2 | | 4918796 | -2.24E+07 | 2 | >60 |
| | 454 25X | 5034500 | 4 | 3 | | 4095943 | -2.24E+07 | 2 | >60 |
| | CCS 22X | 5030885 | 4 | 2 | | 4921886 | -2.24E+07 | 2 | >60 |
| | **Self** | **5029197** | 4 | **2** | | **4919684** | **-2.24E+07** | **2** | **>60** |

**Table 3: Genome assembly continuity and correctness comparison to secondary technologies.** Organism: The genome being assembled. Assembled with: the sequencing data used for assembly. 454 sequencing was unpaired FLX+, with paired-end sequencing available for some genomes, as indicated. MiSeq sequencing was paired-end, indicated as 2xXbp Yb where X is the target read length and Y is the paired-end size. Column definitions are the same as in Table 2. PacBio RS sequences were self-corrected and assembled as in Table 2. 454 Sequences were assembled with Newbler [38] and MiSeq sequences were assembled with SPAdes [51] and MaSuRCA [37, 52]. Both assemblies were polished using iCORN [53] and the one with the best LAP score was reported.

| Organism | Assembled with | Assembly bp | Contigs | N50 | LAP | Discordant Bases | QV |
|---|---|---|---|---|---|---|---|
| *E. coli* K12 | MiSeq 100X 2x150bp 300bp (MaSuRCA iCORN) | 4682345 | 139 | 113852 | 9.68E+07 | 28 | 52.23 |
|  | 454 50X | 4569757 | 93 | 117490 | -9.73E+07 | 17 | 54.29 |
|  | **PBcR 200X** | **4653486** | **1** | **4653486** | **-9.64E+07** | **3** | **>60** |
| *E. coli* O157:H7 | MiSeq 100X 2x150bp 500bp (SPAdes iCORN) | 5433737 | 413 | 133641 | -3.67E+07 | 62 | 49.43 |
|  | 454 22X + 8X 5Kbp + 10X 10Kbp | 5347050 | 409 | 133665 | -3.73E+07 | 66 | 49.09 |
|  | **PBcR 200X** | **5611389** | **9** | **4324437** | **-3.66E+07** | **0** | **>60** |
| *B. trehalosi* | MiSeq 100X 2x150bp 500bp (SPAdes iCORN) | 2377594 | 83 | 222446 | -3.31E+07 | 10 | 53.76 |
|  | 454 50X | 2364704 | 66 | 117742 | -3.32E+07 | 9 | 54.20 |
|  | **PBcR 200X** | **2411068** | **1** | **2411068** | **-3.27E+07** | **0** | **>60** |
| *M. haemolytica* | MiSeq 100X 2x150bp 500bp (MaSuRCA iCORN) | 2721965 | 89 | 84094 | -3.33E+07 | 47 | 47.63 |
|  | **PBcR 200X** | **2736037** | **1** | **2736037** | **-3.31E+07** | **0** | **>60** |
| *F. tularensis* | MiSeq 100X 2x250bp 500bp (SPAdes iCORN) | 1825374 | 130 | 24065 | -1.33E+07 | 0 | >60 |
|  | 454 50X | 1655657 | 326 | 7316 | -1.33E+07 | 28 | 47.72 |
|  | **PBcR 300X** | **1877407** | **3** | **573021** | **-1.33E+07** | **0** | **>60** |
| *S. enterica* Newport | MiSeq 56X 2x150bp 500bp (MaSuRCA iCORN) | 5187269 | 114 | 195780 | -2.24E+07 | 360 | 41.59 |
|  | 454 23X + 2X 10Kbp | 5005089 | 172 | 372513 | -2.25E+07 | 39 | 51.08 |
|  | **PBcR 200X** | **5029197** | **2** | **4919684** | **-2.24E+07** | **2** | **>60** |

**Table 4: *E. coli* correctness on GAGE metrics.** Organism: The genome being assembled. Assembled with: the short-read data used for correction and/or assembly. # Contigs: The number of contigs comprising the assembly. N50: $N$ such that 50% of the genome is contained in contigs of length $\geq N$. # structural differences: the sum of inversions, relocations, and translocations versus the reference using GAGE metrics [36]. # Discordant bases: total number of different bases when compared to the reference. QV: estimated from the number of indels and SNPs as $\log_{10}\left(\frac{\text{assembly length}}{\text{\# incorrect bases}}\right) * 10$. Indels > 5bp: The number of indels greater than 5bp reported by GAGE metrics [36]. Assemblies were generated as in Tables 2 and 3.

| Organism | Assembled By | # Contigs | N50 | # Structural differences | # Discordant Bases | QV | # Indels > 5bp |
|---|---|---|---|---|---|---|---|
| *E. coli* K12 | MiSeq 100X | 139 | 113852 | 1 | 59 | 49.00 | 4 |
| | 454 50X | 93 | 117490 | 2 | 26 | 52.45 | 0 |
| | MiSeq 100X + 200X CLR | 2 | 2367319 | 2 | 11 | 56.26 | 2 |
| | 454 50X + 200X CLR | 1 | 4649004 | 2 | 11 | 56.26 | 2 |
| | CCS 25X + 200X CLR | 1 | 4653267 | 0 | 14 | 55.22 | 2 |
| | 200X CLR | 1 | 4653486 | 0 | 14 | 55.22 | 2 |